\documentstyle[12pt,openbib]{article}
\def\be{\begin{equation}}           
\def\ee{\end{equation}}

\newcommand{\La}{\Lambda}

\newcommand{\ka}{\kappa}
\newcommand{\s}{\Sigma}
\newcommand{\ca}{\Xi}
\newcommand{\g}{\gamma}
\newcommand{\Om}{\Omega}
\newcommand{\r}{\rho}
\newcommand{\f}{\frac}

\newcommand{\ep}{\epsilon}
\begin{document}
\title{\bf Magnetic Moment of the $\Om ^-$ in QCD sumrule (QCDSR)}
\vskip .5cm
\author{Jishnu Dey $^{1,3}$, Mira Dey  $^{2,3}$, 
 \\ and \\Ashik Iqubal $^{4}$}
\vspace{.5 cm}
\date{\today }
\maketitle
{\it Abstract} :
The $\Om ^-$ magnetic moment was measured very accurately and experimentalists remarked 
that it differs from the theoretical estimates, thus posing a challenge to the latter. 
One such estimation uses QCDSR. We revisit this sumrule method, using condensate 
parameters
which were obtained from fitting the differences ($\mu _p -  \mu _n$), ($\mu _{\s^+} -
\mu _{\s^-}$) and ($\mu _{\ca^0} - \mu _{\ca^-}$) \cite{ddsr} and confirm the 
experimental 
number. The $\mu_{\Delta^{++}}$ is also found to agree with the experimental 
estimate.  
\vskip .5cm

(1) Azad Physics Centre, Dept. of Physics, Maulana Azad College, Calcutta 700
013, India and ICTP, Trieste, Italy\\ 
(2) Dept. of Physics, Presidency College, Calcutta 700 073,
India and and ICTP, Trieste, Italy \\
(3) {\it{ Work supported in part by DST grant no. SP/S2/K18/96, Govt.
of India, \\ permanent address : 1/10 Prince Golam Md. Road,
Calcutta 700 026, India, email : deyjm@giascl01.vsnl.net.in}}.\\
(4) University College of Science, University of Calcutta, Acharya Prafulla
Chandra Road, Calcutta, India.
\newpage

The $\Om ^-$ magnetic moment, $\mu _{\Om^-}$, has been the subject of many studies 
\cite{ddt,lee,Latt,Rqm,CSqm}. The magnetic moment was unknown when \cite{ddt} was 
published but on hindsight the value predicted there, within the acceptable parameter 
range, 
agrees with the present accurately determined  experimental result \cite{wall}.  The 
results 
of Lee \cite{lee} using 
QCD sumrules and those from the lattice calculation \cite{Latt} underestimate it 
whereas the 
light- cone relativistic quark model \cite{Rqm} and the chiral quark soliton model 
\cite{CSqm} overestimate it. We re-investigate this intriguing situation by looking at 
the calculations of Lee using a slightly different point of view advocated in 
\cite{ddsr} 
and find that one indeed gets good agreement with experiment. Further, as pointed out 
by Lee, 
the  $\mu _{\Om^-}$ depend sensitively on the magnetic susceptibility so that we can 
pinpoint this parameter more effectively. 

The QCD sumrule method is a very powerful tool in revealing a deep connection between 
hadron 
phenomenology and vacuum structure \cite{svz} via a few condensates like 
\be
a = - 2 \pi^2<\bar q q>, \;b = <g^2 G^2>
\label{eq:1}, 
\ee
related to the quark (q) and gluon (G) vacuum expectation values.
This can be used for evaluating magnetic moments of hadrons \cite{si} where some new 
parameters enter, for example, $\chi$, $\ka$ and $\xi$, defined through the following 
equations :
\be
<\bar q\sigma_{\mu\nu}q>_F = e_q \chi <\bar q q> F_{\mu \nu}
\label{eq:2}, 
\ee
\be
<\bar q g G_{\mu\nu}q>_F = e_q \ka <\bar q q> F_{\mu \nu}
\label{eq:3},
\ee
\be
<\bar q\ep_{\mu\nu\r\g}G^{\r\g}\g_5q> = e_q \xi <\bar q q> F_{\mu \nu}
\label{eq:4}. 
\ee
where the F denotes the usual external electromagnetic field tensor. Lee \cite{lee} 
very 
carefully evaluated the contributions of these operators to the magnetic moments of the 
$\Om ^-$ and $\Delta^{++}$, the latter emerging from the former when the quark mass 
$m_s$, 
is put equal to zero, the parameter $f$ and $\phi$ are put equal to 1 and the quark 
charge $e_s = -1/3$ is replaced by $e_u = 2/3$. The parameter $f$ and $\phi$ measure 
the values of quark condensates and quark spin-condensates with strange and (ud) 
quarks.  
\be
f = \f{<\bar s s>}{\bar u u}
\label{eq:5}, 
\ee  
\be
\phi = \f{<\bar s\sigma_{\mu\nu}s>}{<\bar u\sigma_{\mu\nu}u> } 
\label{eq:6}
\ee

For the expression for the  $\mu _{\Om^-}$ sumrule we refer the reader to the paper by 
Lee \cite{lee} which we reproduce here for the sake of completeness, in terms of the 
Borel parameter M and the intermediate state contribution A  :
 
\begin{eqnarray}
& &
  {9\over 28} e_s  L^{4/27} E_1 M^4
- {15\over 7} e_s f \phi  m_s \chi a L^{-12/27} E_0 M^2
+ {3\over 56} e_s  b L^{4/27} 
- {18\over 7} e_s f m_s a L^{4/27} 
\nonumber \\ & &
- {9\over 28} e_s f \phi (2\ka + \xi) m_s a L^{4/27} 
- {6\over 7} e_s f^2 \phi \chi a^2 L^{12/27}
- {4\over 7} e_s f^2 \ka_v a^2 L^{28/27} {1\over M^2}
\nonumber \\ & &
- {1\over 14} e_s f^2 \phi (4\ka + \xi) a^2 L^{28/27} {1\over M^2}
+ {1\over 4} e_s f^2 \phi \chi m^2_0 a^2 L^{-2/27} {1\over M^2}
\nonumber \\ & &
- {9\over 28} e_s f m_s m^2_0 a L^{-10/27} {1\over M^2}
+ {1\over 12} e_s f^2 m^2_0 a^2 L^{14/27} {1\over M^4}
\nonumber \\ & &
= \tilde{\lambda}^2_\Om
\left( {\mu_{\scriptscriptstyle \Om}\over M^2} + A \right) e^{-M^2_\Om/M^2}.
\label{omeg_we5}
\end{eqnarray}
Here
\be
E_n(x) = 1 - e^{-x}\sum_
n \f{x^n}{n!},\; x = w_B^2/M_B^2
\ee
where $w_B$ is the continuum, and 
\be
L = \f{ln(M^2/\La ^2_{QCD})}{ln(\mu ^2/\La ^2_{QCD})}  
\ee
For evaluating the magnetic moment we use the above equation and divide by the equation 
for 
the mass sumrule given earlier by Lee \cite{lee2}. Thus we eliminate the parameter 
$\lambda_\Om ^-$  in the spirit of \cite{ddsr} and we get an excellent fit to the 
resulting numbers in the form 
$\mu _{\Om^-} + A/M^2$. We find that the results are not very sensitive to $\ka_v$, the 
so called factorization violation parameter, defined through 
\be
<\bar u u \bar u u> = \ka_v <\bar u>^2.
\ee
Neither are the results very sensitive to 
the parameters $\ka$ and $\xi$. We use the crucial parameters $a$ and $b$ from 
\cite{ddsr}, since they must fit the octet baryon moment-differences ($\mu _p - 
\mu _n$) and ($\mu _{\s^+} - \mu _{\s^-}$). It was shown in \cite{ddsr} 
that by using the empirical scaling of the  $\tilde{\lambda}$ 
with the $(baryon\; mass)^3$ - these differences depend only of $a$ and $b$,
and one gets $a = 0.475\; GeV^3$ and $b = 1.695 \;GeV^4$. Further, to fit the 
difference  
$(\mu _{\ca^0} - \mu _{\ca^-})$, $m_s$ was set to be 170 MeV in \cite{ddsr} and we 
use this value.     

Table 1 shows the dependence of the magnetic moments on the parameters. Obviously
$\mu _{\Delta^{++}}$ does not depend on $f$ and $\phi$. It is clear that 
$\mu _{\Om^-}$ also does not depend so much on $f$ but it is sensitive to both
$\phi$ and $\chi$, and it appears that ($\chi \,=\,6.5, \phi = 0.6$) and 
($\chi \,=\,5.5, \phi = 0.7$) are preferred values, close to the experimental number
$\mu _{\Om^-}\, =\, 2.019 \pm 0.054\,\mu _N $ \cite{wall}. The $\mu _{\Delta^
{++}}$ is known only approximately, $4.52 \pm 0.95\, \mu _N$ \cite{boss} and a better 
determination will enable us to pinpoint $\chi$.

It is satisfactory to see that there is no conflict between experiment and QCDSR since 
sum rules are a `first principle method', although it is based partly on phenomenology.

In summary we find that using the constrained values of the parameters $a$ and $b$ 
\cite{ddsr} one can get a good fit to the known decuplet magnetic moments. The moments
do not depend sensitively on the factroization violation parameter but may be used to 
pinpoint the susceptibility $\chi$ and $\phi$, the ratio of the spin condensate for strange and ud quarks.

\vskip 2.5cm 
\noindent{\small {\bf Table 1.} \ The values of the parameters and the 
corresponding magnetic moments. 
\vskip 1.5cm
\begin{center}
\begin{tabular}{|c|c|c|c|c|c|c|c|}
\hline
$\ka$&$\xi$&$\chi$&$\ka_v$ &$f$&$\phi$&$\mu _{\Om^-}$&$\mu _{\Delta^
{++}}$\\
\hline
0.70 & -1.5 & -6.5 &1.0&0.83 &0.6 & -2.007&3.702\\
0.75 & -1.5 & -6.5 &1.0&0.83 &0.6 & -2.005&3.697 \\
0.80 & -1.5 & -6.5 &1.0&0.83 &0.6 & -2.002&3.691\\
0.75 & -1.4 & -6.5 &1.0&0.83 &0.6 & -1.983&3.670 \\
0.75 & -1.6 & -6.5 &1.0&0.83 &0.6 & -2.026&3.724 \\
0.75 & -1.5 & -7.0 &1.0&0.83 &0.6 & -2.146&3.964 \\
0.75 & -1.5 & -6.0 &1.0&0.83 &0.6 & -1.884&3.457 \\
0.75 & -1.5 & -6.5 &1.5&0.83 &0.6 & -1.928&3.588 \\
0.75 & -1.5 & -6.5 &1.0&0.83 &0.7 & -2.750&3.697 \\
0.75 & -1.5 & -5.5 &1.0&0.83 &0.7 & -2.011&3.217 \\
0.75 & -1.5 & -6.5 &1.0&0.88 &0.6 & -2.020&3.697 \\

\hline 
\end{tabular}
\end{center}

\end{document}